# Ring Oscillator Physical Unclonable Function with Multi Level Supply Voltages

Shohreh Sharif Mansouri and Elena Dubrova
Department of Electronic Systems, School of ICT, KTH - Royal Institute of Technology, Stockholm
Email:{shsm,dubrova}@kth.se

*Abstract*—One of the most common types of Physical Unclonable Functions (PUFs) is the ring oscillator PUF (RO-PUF), in which the output bits are obtained by comparing the oscillation frequencies of different ring oscillators. In this paper we design a new type of ring oscillator PUF in which the different inverters composing the ring oscillators can be supplied by different voltages. The new RO-PUF can be used to (1) increase the maximum number of possible challenge/response pairs produced by the PUF; (2) generate a high number of bits while consuming a low area; (3) improve the reliability of the PUF in case of temperature variations. We present the basic idea of the new RO-PUF and then discuss its applications.

## I. INTRODUCTION

Most secure cryptographic algorithms use a private secret value, defined as *key*, to encrypt and decrypt data. The key is normally stored in a RAM or a non-volatile memory. Being dependent to a unique value stored in memory makes algorithms vulnerable to various attacks such as invasive and semi-invasive attacks like physical tampering [1]–[3], whose goal is to obtain access to the key. Protecting the stored key against these attacks is vital for guaranteeing the security of the cryptographic devices.

As suggested in [2], possible techniques used during manufacturing, such as using fuse memory arrays and planarising each predecessor layer before applying the next layer, can partially protect the storage elements against these attacks. However, the attackers are constantly looking for new attack methods. There is a non-stop ongoing battle between the designers who are trying to improve the security of their products and the attackers who are constantly trying to break them [2].

Physical Unclonable Functions (PUFs) were introduced in 2002 [4] by B. Gassend and co-workers. PUFs use the physical structure of each device to generate a set of unique data which resembles the chip fingerprint. Even if identical PUFs are implemented in different chips using the same manufacturing process, small device-to-device variations result in each PUF generating a different set of data, which is in principle unique and impossible to duplicate for all chips. PUFs are sensitive to device variations; any invasive or semi-invasive attack will, with a high probability, cause a permanent alteration of the device physical properties and thus alter permanently the behaviour of the PUF, rendering the device unusable. Therefore, it is believed that Physical Unclonable Functions provide high security for hardware devices. Two main usages of PUFs in crypto-systems are embodying a single cryptographic key and implementing a challenge-response authentication method.

Ring Oscillator PUFs (RO-PUF) were introduced in 2007 [5] and exploit the differences between the delay characteristics of wires and transistors. The output bits of a RO-PUF are determined by comparing the oscillation frequencies of ring oscillators. RO-PUFs have a high reliability and are easier to implement compared to previously proposed designs such as butterfly PUFs [6]. Since 2007, many researches were conducted on RO-PUFs. In [7] and [8], a possible implementation of a RO-PUF on an FPGA was suggested. [9] and [10] introduced methods to increase the reliability in case of temperature variations; other works aimed at making the hardware more secure [11].

In this work we suggest a new design for RO-PUFs, which is based on the idea to use independent supply voltages for the different inverters composing all ring oscillators. Our method can be used for:

- Improving authentication by increasing the maximum number of possible challenge/response pairs produced by a RO-PUF.
- Designing low-area RO-PUFs for applications with strong area limitations
- Improving the RO-PUF reliability by decreasing its sensitivity to temperature variations

All three applications are analysed in detail, comparing the results we obtain with state-of-art solutions that can be found in literature.

The remainder of the paper is organized as follows: in Section II, an overview of RO-PUFs is given; in Section III we discuss the dependency of ring oscillator frequencies to supply voltage variations; in Section IV we introduce the basic idea of our new RO-PUF; in Section V we calculate the uniqueness of our RO-PUF; in Section VI we study how our RO-PUF can be used to implement an efficient challenge-response authentication method; in Section VII we estimate the area savings that can be obtained using our RO-PUF; in Section VIII we suggest how our RO-PUF can be used to increase reliability in presence of temperature variations; in Section IX we conclude the paper and discuss future works.

## II. RING OSCILLATOR PUFS

Ring Oscillator PUFs (RO-PUFs), as designed in [5], have a simple architecture made of two n-bit multiplexer, 2 counters, 1 comparator and $n$ ring oscillators (ROs) (see Figure 1).

Each ring oscillator contains an odd number of inverters connected in a loop; each ring oscillates with a unique frequency depending on the characteristics of each of its inverters, which variate unpredictably from cell to cell due to manufacturing variations, even within the same chip, and are impossible to imitate. If the frequencies at which the ring oscillators oscillate are too high, the counters may not be able to count oscillations; therefore, there is a minimal number of inverters in every ring oscillator necessary to ensure a suitable oscillating frequency. This value depends on the technology but is typically in the order of 10 - 20 inverters.

The two multiplexers select two ROs which are compared together (pair). The two counter blocks count the number of oscillations of each of the two ROs in a fixed time interval (comparison time). At the end of the interval, the outputs of the two counters are compared together. Depending on which of the two counters has the highest value, the output of the PUF is set to 0 or 1. The output of the PUF is set to 0 if the first ring oscillator in the pair is faster than the second (the value of the first counter is higher than that of the second), and to 1 if it is slower (the value of the first counter is lower than that of the second). If the two frequencies are very close to each other, the output of the PUF may variate unpredictably from run to run. It is however possible to improve the accuracy of the PUF by using larger counters and longer comparison time intervals.

Originally, a RO-PUF produces 1-bit output for each comparison time interval. In each comparison time interval, the multiplexer selector is changed, the pair is changed and the RO-PUF produces another bit. A RO-PUF can be modified to produce multiple bits of data per comparison interval by increasing the number of multiplexers, counters and comparators, and comparing several pairs of ring oscillators at the same time.

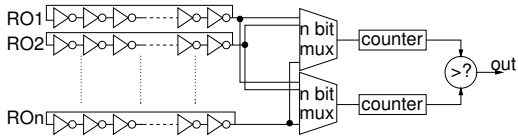

Fig. 1. A RO-PUF circuit implemented as in [5]

### III. VOLTAGE SUPPLY DEPENDENCY

All electronic components are sensitive to variations of their operating conditions. Supply voltage variations in CMOS systems affect the delay $d$ of a combinational path of digital cells: if the supply voltage raises, the delay decreases and vice-versa. The relation between the delay $d$ and the supply voltage $V$ of the path is a complex relation which can be measured experimentally or estimated at SPICE level. A simplified, approximate model for this relation is given by the Alpha law [12]:

$$d = K \frac{V}{(V - V_{th})^\alpha} d_M \qquad (1)$$

Where $K = \frac{(V_M - V_{th})^\alpha}{V_M}$ is a scaling factor; $V_{th}$ is the threshold voltage of the transistors, $\alpha$ is the velocity saturation index of the technology and $d_M$ is the delay of the path when supplied with the typical supply voltage $V_M$ of the technology. In UMC 90 nm ASIC technology, $V_{th} = 0.6V$, $\alpha = 1.54$ and $K = 0.379$ were estimated by simulating at SPICE level a chain of 10 inverters under different supply voltages. In principle the Alpha law is valid for all combinational paths in a given technology and the only parameter that changes from path to path is the scaling factor $d_M$.

### IV. MULTI-VOLTAGE RO-PUF

Our idea is to exploit the delay-to-supply voltage dependency from formula 1 to transform an original RO-PUF into a new RO-PUF characterized by smaller design, higher number of bits and higher resistance to temperature variations.

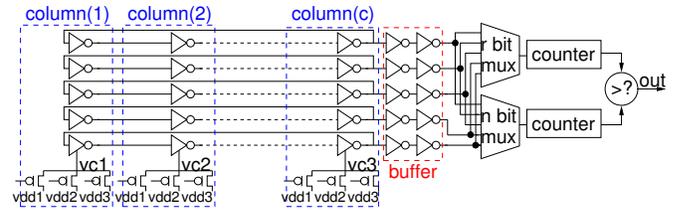

Fig. 2. Multi-voltage RO-PUF

The basic scheme of our RO-PUF is shown in Figure 2. In all ring oscillators, the inverters are grouped in $C$ different columns with roughly the same number of inverters. Inverters belonging to the same column always share the same supply voltage. The supply voltage of each column of inverters can be selected among $L$ different values (In Figure 2 $L = 3$) indicated as $Vdd_1, Vdd_2, ..., Vdd_L$ and can be selected independently from the supply voltage of the other columns using PMOS switches. The columns can contain a single inverter per ring or multiple inverters grouped together. The supply voltages $Vdd_1, ..., Vdd_L$ are selected within the technology range allowing direct communication between gates without requiring level shifters, so that inverters can communicate with each other safely. The multiplexers, the counters and the comparators operate always on typical supply voltage. Buffers (two inverters in series operating under typical voltage supply) are introduced at the output of every ring oscillator to guarantee fast signal transitions at the input of the multiplexers.

The oscillation frequency of each RO depends on the delay of all the inverters composing it. As an example, let us consider the UMC 90 nm PUF in Figure 3, where two ring oscillators indicated as $RO_A$ and $RO_B$ are being compared. Both ring oscillators are composed of 3 inverters, with $C = 3$ and $L = 3$ (small ring oscillators are chosen to keep the presentation simple, but note that these ring oscillators are too fast to be used in a real PUF). For the case in which all column supply voltages are equal to the typical supply voltage of the technology $Vdd_2 = 1.2V$, the relation between the delays of the inverters $d_i$ and the total delay $d_{RO}$ of the ROs (the inverse of their oscillating frequency) can be defined as:

$$d_{i1A} + d_{i2A} + d_{i3A} = d_{ROA}$$
$$d_{i1B} + d_{i2B} + d_{i3B} = d_{ROB} \quad (2)$$

In general, $d_{i1}$, $d_{i2}$ and $d_{i3}$ are not equal and depend on unpredictable and uncontrollable device variations between the different inverters composing each ring oscillator. As an example, we suppose to have:

$$3d_{i1A} = 2d_{i2A} = 6d_{i3A}$$
$$4d_{i1B} = 4d_{i2B} = d_{i3B} \quad (3)$$

Combining relations 2 and 3, the components of $d_{RO}$ are given by:

$$\frac{2}{6}d_{ROA} + \frac{3}{6}d_{ROA} + \frac{1}{6}d_{ROA} = d_{ROA}$$
$$\frac{1}{6}d_{ROB} + \frac{1}{6}d_{ROB} + \frac{4}{6}d_{ROB} = d_{ROB}$$

As discussed in Section III, the supply voltage variation of each electronic component directly affects its delay based on relation 1. If the supply voltage of the first inverter in both ring oscillators is raised from $1.2V$ to $1.32V$, the relation predicts that its delay will decrease by a factor $\sim 0.83$. Based on formula 1, the delays of the ring oscillators $d^*_{RO}$ under the new configuration convert to:

$$d^*_{ROA} = 0.83\frac{2}{6}d_{ROA} + \frac{3}{6}d_{ROA} + \frac{1}{6}d_{ROA} = 0.94d_{ROA}$$
$$d^*_{ROB} = 0.83\frac{1}{6}d_{ROB} + \frac{1}{6}d_{ROB} + \frac{4}{6}d_{ROB} = 0.97d_{ROB}$$

Due to the increase in the speed of the first inverter in both ROs, the new oscillating frequencies $f^*_{ROA} = \frac{1}{d^*_{ROA}}$ and $f^*_{ROB} = \frac{1}{d^*_{ROB}}$ are higher than the nominal frequencies $f_{ROA} = \frac{1}{d_{ROA}}$ and $f_{ROB} = \frac{1}{d_{ROB}}$. However, since $f^*_{ROA} = \frac{f_{ROA}}{0.94}$ and $f^*_{ROB} = \frac{f_{ROB}}{0.97}$, there is no guarantee that these frequencies hold the same relation as they did under nominal supply voltage. In other words, if $RO_A$ was slower than $RO_B$ under nominal supply voltages, there is a possibility that $RO_A$ is faster than $RO_B$ with the new supply voltage configuration.

In our RO-PUF, instead of a single bit output, by changing the supply voltages of the different columns, each pair of ring oscillators can produce a set of different output bits. For each pair, the maximum number of bits which can be produced with $C$ columns and $L$ supply voltages is equal to $L^C$. As an example, for a RO-PUF with $L = 3$ and $C = 3$, the number of bits that can be generated by one pair of ring oscillators is $3^3 = 27$.

We generated using SPICE two random 3-inverter ring oscillators in UMC 90 nm technology with $C = 3$ and $L = 3$ ($Vdd_1 = 1.08V$; $Vdd_2 = 1.2V$; $Vdd_3 = 1.32V$), setting the characteristics of each device randomly depending on the typical manufacturing intra-chip tolerances of the technology. We then set the supply voltages of every column to all possible $L^C = 27$ configurations. The relation between the delays of $RO_A$ and $RO_B$ ($d_{ROA} - d_{ROB}$) for all these configurations is shown in Figure 4. Among all the configurations, 12 result in the first RO being faster than the second, and the other 15 result in the second RO being faster than the first.

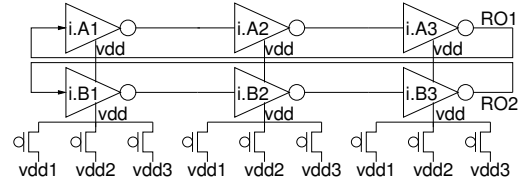

Fig. 3. A pair of three-inverter ring oscillators with $C = 3$ and $L = 3$.

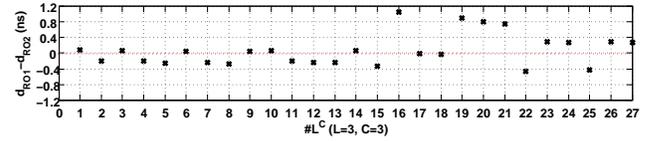

Fig. 4. Values of $d_{ROA} - d_{ROB}$ for all 27 voltage configurations for the RO pair in Figure 3. $d_{ROA}$ and $d_{ROB}$ indicate respectively the delays of $RO_A$ and $RO_B$. If $RO_A$ is faster than $RO_B$, $d_{ROA} - d_{ROB}$ is negative, else it is positive.

This increase in output bits cannot be obtained by controlling the supply voltage of the entire PUF instead of controlling independently the supply voltage of each inverter column. In fact, if the supply voltage of a whole PUF is changed, then both ring oscillators composing any pair will work at a higher or lower supply voltage. Based on the Alpha law no variation in the output bit of the PUF should occur, because both ring oscillators in the pair operate faster or slower with ideally matched variations.

Continuing with the same example RO-PUF from Figure 3 and defined by equations 2 and 3, if all gates work with the highest supply voltage $Vdd_3 = 1.32V$, the delays of the ring oscillators $d^+_{RO}$ under the new configuration convert to:

$$d^+_{ROA} = 0.83\frac{2}{6}d_{ROA} + 0.83\frac{3}{6}d_{ROA} + 0.83\frac{1}{6}d_{ROA} = 0.83d_{ROA}$$
$$d^+_{ROB} = 0.83\frac{1}{6}d_{ROB} + 0.83\frac{1}{6}d_{ROB} + 0.83\frac{4}{6}d_{ROB} = 0.83d_{ROB}$$

Since the variations have the same effect on all six inverters, they have the same effect on the delays on the ring oscillators and the RO-PUF output bit does not change. Experimental results reported in [5] are consistent with this analysis and show that by changing the supply voltage of a RO-PUF by $10\%$ from the typical voltage, only $0.48\%$ of the bits flip their value. The bits that flip are explained by higher order effects which are not considered in the Alpha law and which affect mainly pairs of ring oscillators that run at closely matched frequencies.

Our solution constrains all inverters in the same column to always be supplied by the same supply voltage. The idea of a

RO-PUF is to compare the oscillating frequencies of ring oscillators that are nominally identical, and without this constraint this would not be the case. An attacker having gained access to the supply voltage configuration of a chip through an invasive or semi-invasive attack can gain knowledge on the structure of the PUF and thus guess the most probable output bit values. Also, the attacker could modify the supply voltages so that one of the output bits is changed in a predictable fashion.

Compared to the original RO-PUF, we support multiple supply voltages in our design. Our design should be reliable in case of voltage variations in all the supply levels. To guarantee that the system operates reliably, is that the voltage distance between each two supply levels should satisfy the relation:

$$MAX(|Vdd_i - Vdd_j|) > \frac{MAX(VAR_i + VAR_j)}{2}$$
$$(1 < i, j \leq L) \quad (4)$$

where $VAR_i$ and $VAR_j$ are defined respectively as the maximum variations of $Vdd_i$ and $Vdd_j$ (see Figure 5).

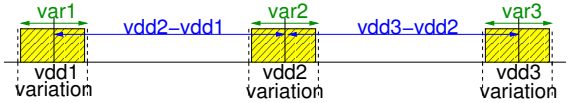

Fig. 5. Voltage levels contraints

## V. UNIQUENESS

Inter-chip uniqueness is a parameter typically used to evaluate PUFs. The inter-chip uniqueness of a PUF is calculated by checking how different and random are the output bits of two identical PUFs implemented in different chips.

As discussed in [13], the uniqueness $U$ for an original RO-PUF can be calculated by considering a set of $k$ identical PUFs implemented in different chips. The uniqueness is defined as the average Hamming Distance between $n$ bit outputs obtained from any possible pair of two PUFs $i$ and $j$, expressed in percentage:

$$U = \binom{k}{2}^{-1} \sum_{i=1}^{i=k-1} \sum_{j=i+1}^{j=k} \frac{HD\left(O_i(n), O_j(n)\right)}{n} \times 100\%$$
$$= \frac{2}{k \times (k-1)} \sum_{i=1}^{i=k-1} \sum_{j=i+1}^{j=k} \frac{HD\left(O_i(n), O_j(n)\right)}{n} \times 100\%$$

where $HD\left(O_i(n), O_j(n)\right)$ is the Hamming Distance between two series of n-bit outputs ($O_i(n)$ and $O_j(n)$) obtained by setting the multiplexers of the two PUFs to $n$ different values (identical for PUFs $i$ and $j$) during $n$ different comparison intervals.

With our RO-PUF, the output bits of a PUF depend not only on which ring oscillators are selected by the multiplexers but also on the supply voltage configuration. The uniqueness is estimated with the same formula as for the original RO-PUF, but each of the $n$ output bits is obtained by setting the multiplexers and the voltage supplies to $n$ different configurations (identical for PUFs $i$ and $j$) during $n$ different comparison intervals.

We designed $k = 20$ RO-PUFs in UMC 90 nm technology, each containing only two ring oscillators composed of 13 inverters, with $C = 3$ and $L = 2$. The RO-PUFs are all nominally identical, but all devices where generated with random characteristic mismatches based on the typical manufacturing tolerances of the technology. Through SPICE simulation, with $n = L^C = 8$ we found $U = 51.35\%$, which is close to the ideal result 50%.

## VI. AUTHENTICATION

As discussed in [5], RO-PUFs can be used to implement challenge-response authentication protocols. In the original RO-PUF, such as the one shown in Figure 2, the multiplexer selector bits that define which ring oscillators should be paired together are used as challenge bits, i.e. they are set by the system with which the PUF-enabled device communicates, which knows what output it should expect from the cryptographic device. The response to the challenge is defined by the device based on the frequency comparison of the two selected ROs, and checked against tables of expected responses.

For a traditional RO-PUF composed of $R$ ring oscillators, the maximum number of challenges is given by $\frac{R(R-1)}{2}$, which corresponds to the number of possible ring oscillator pairs.

With our RO-PUF, shown in Figure 2, the challenge bits can be set not only by changing the selectors of the multiplexers, but also by setting the voltage configuration of the PUF, i.e. the supply voltage of every column of inverters. The maximum number of challenge/response pairs is increased to

$$\frac{R(R-1)}{2} \times L^C$$

where $R$ is number of ring oscillators, $C$ is the number of columns and $L$ is the number of supply voltages.

As pointed out in [5], for the original RO-PUF not all challenges are valid: if, for example, $RO_A$ is faster than $RO_B$ and $RO_B$ is faster than $RO_C$, then $RO_A$ is necessarily faster than $RO_C$. The result from the challenge selecting ring oscillators $RO_A$ and $RO_C$ is predictable if the responses to the other two challenges are known and does not constitute a valid challenge. Also, as discussed in [5], RO-PUF circuits are sensitive to temperature variations: under certain challenges, the output bit of the PUF may flip depending on the temperature. Only challenges that do not exhibit this behaviour are valid.

Similarly, in our RO-PUF not all challenges will be valid. Two voltage configurations which are very near to each other (with only a very limited number of supply voltages changing between the two) have a high probability to result in the same output bit. We leave a precise determination of the number of valid challenge-response pairs to future works. In this paper, we compare our RO-PUF with the original RO-PUF only in terms of total number of challenge-response pairs, without considering the impact of the invalid pairs.

## VII. AREA CONSIDERATIONS

Many systems using cryptographic algorithms such as hardware authentication devices (RFID tags, etc.), smartcards, and wireless networks (Bluetooth, NFC, tags, etc.) are characterized by very tight power and area budgets. One of the main advantages of our RO-PUF is its ability to produce the same number of bits than the original RO-PUF [5] using a smaller hardware.

In the original RO-PUF, the maximum number of bits produced by the PUF can be increased only by adding more ring oscillators to the PUF (see Figure 6-B). Since the structure of a RO-PUF is simple, the ring oscillators make up most of the RO-PUF area, and increasing the number of ring oscillators easily increases its total area.

With our RO-PUF, the maximum number of bits depends not only to number of ring oscillators but also on the number of columns and supply voltages. Increasing each of these three elements increases the number of bits produced by the PUF. From Figure 6, it is obvious that increasing the number of columns instead of the number of ring oscillators is a more effective way to increase the output bits of the PUF.

By increasing the number of columns in each PUF without adding to the number of ring oscillators, we achieve the same improvement in the number of bits that can be generated by a traditional RO-PUF, but with a smaller hardware.

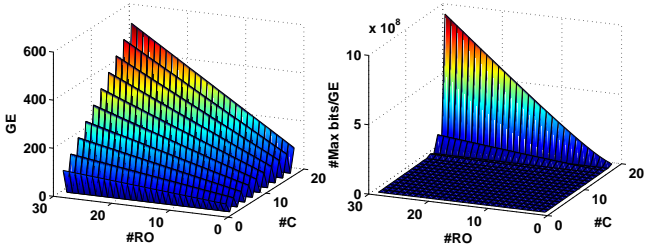

Fig. 7. Area (left) and maximum number of output bits per unit of area (right) for a series of our RO-PUFs with $L = 3$, $2 \leq R \leq 30$, number of inverters per RO between 3 and 19 (odd values only) and $C$ equal to the number of inverters per RO.

Figure 7-left shows the area in terms of gate count for a set of UMC 90 nm RO-PUFs with different number of ROs and columns.

Brought to the extreme, this leads to what is to our best knowledge the most compact RO-PUF suggested in literature: as shown in Figure 8 the RO-PUF is made of only two ring oscillators which act as a pair, and there are no multiplexers. Table I reports implementation details of the RO-PUF in Figure 8 implemented in UMC 90 nm technology. Different values of $L$ and $C$ are considered. The number of inverters is kept equal to $C$ (one inverter per column per ring oscillator). Three values of $L$ and $C$ are chosen to have $L^C > 2^{22}$, $L^C > 2^{80}$ and $L^C > 2^{160}$. Frequency results represent the oscillating frequency of the ring oscillators under typical supply voltage conditions obtained from Cadence RTL Compiler; power results are estimated as a combination of dynamic and leakage power with a typical supply voltage, at $160MHz$ clock frequency. The areas of the power switches are not considered in Table I.

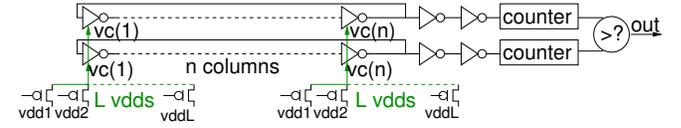

Fig. 8. A low-area RO-PUF circuit.

The area overhead of the PMOS power switches is the only overhead which is added to the original RO-PUF. The percentage of this overhead on the original RO-PUF is shown in Figure 9. The highest overhead (42%) is for a PUF with only 1 pair of ring oscillators, 19 inverters per RO and $C = 19$. However, this RO-PUF has a much lower area compared to the original RO-PUF which produces the same number of bits (see Figure 7).

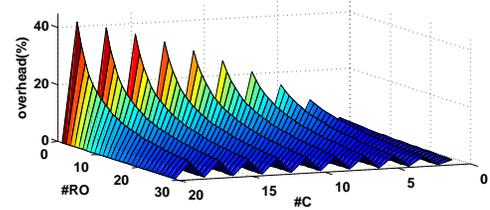

Fig. 9. Area overheads of the supply voltage switches. Each PMOS switch is assumed to consume $0.5GEs$.

## VIII. TEMPERATURE-RESISTANT RO-PUF

All electronic components are sensitive to variations of their operating conditions. For a RO-PUF, this sensitivity can cause uncertainty in the output bits (low reliability). All PUF circuits should be modelled and tested extensively before a PUF is commercially deployed, to guarantee high reliability, i.e. that the cryptographic key or the unique identifier derived from them is exactly the same under all circumstances.

Temperature is one of the main operating conditions that can impact reliability [14]: increasing or decreasing the temperature of a RO-PUF can make some of its output bits flip due to the unequal effect of temperature variations on the two ring oscillators that are being compared. Recently, several temperature-aware RO-PUFs have been suggested in literature.

In [9], the authors suggest to introduce a temperature sensor in every chip. After a chip containing the PUF has been manufactured, it is tested to determine in which temperature range every pair of ring oscillators is reliable. All ring oscillator pairs are used only in the intervals in which they are reliable; an output bit is generated using different pairs depending on the temperature of the PUF: if the temperature changes, the pair of ring oscillators generating a bit is changed. A table coupling temperature ranges with reliable pairs is stored in a memory. Even if an attacker can gain access to the contents of this memory, no information about the PUF structure is revealed. The number of output bits is lower than the total number of

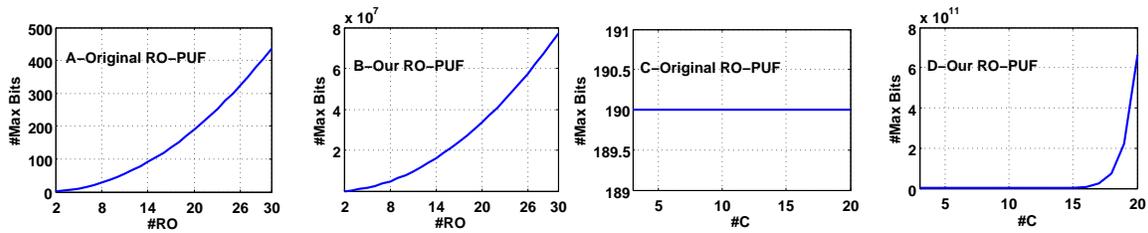

Fig. 6. Comparison between the maximum number of output bits produced by our PUF (B, D) and the original RO-PUF (A, C). A: original RO-PUF with 11 inverters per RO and $2 \leq R \leq 30$; B: our RO-PUF with 11 inverters per RO, $C = 11$, $L = 3$ and $2 \leq R \leq 30$; C: original RO-PUF with $R = 20$ and number of inverters per RO between 3 and 19 (odd values only); D: our RO-PUF with $R = 20$, $L = 3$, number of inverters per RO between 3 and 19 (odd values only) and $C$ equal to the number of inverters per RO.

| #Max bits | $2^{22}$ | | | $2^{80}$ | | | $2^{160}$ | | |
|---|---|---|---|---|---|---|---|---|---|
| $L$ | 2 | 3 | 4 | 2 | 3 | 4 | 2 | 3 | 4 |
| $C$ | 23 | 15 | 11 | 81 | 51 | 41 | 161 | 101 | 81 |
| Area ($\mu m^2$) | 289 | 241 | 223 | 625 | 463 | 397 | 1117 | 763 | 637 |
| Freq ( MHz) | 1141 | 1712 | 2107 | 344.8 | 516.9 | 652.3 | 169.2 | 266.0 | 344.8 |
| Power ($\mu W$) | 81.1 | 60.1 | 52.3 | 233.1 | 157.2 | 128.4 | 451.1 | 288.4 | 233.3 |

TABLE I

AREA, FREQUENCY AND POWER FIGURES OF THE PUF IN FIGURE 8 FOR DIFFERENT VALUES OF $C$ AND $L$ (ONE INVERTER PER COLUMN PER RING OSCILLATOR).

ring oscillator pairs; a hardware utilization of in average 80% can be achieved by this design compared to the original PUF.

The work in [10] is based on the idea that the effects of temperature changes on ring oscillator frequencies can be partially compensated by changing the supply voltage of a PUF. The authors define and store in memory a table which matches each temperature to a corresponding supply voltage. During operation, the temperature is estimated by an on-chip temperature sensor and the supply voltage of the PUF is changed accordingly. The information related to the operational temperatures and the corresponding supply voltages are saved on an on-chip memory. Attacking this memory does not reveal any useful information to the attacker. This work increases reliability but does not guarantee that all pairs of ring oscillators will result in a reliable output bit: some pairs of ring oscillators will still need to be eliminated for the sake of reliability.

With some changes, our multi-level RO-PUF can be transformed into a temperature-aware RO-PUF that can cope with temperature variations. The main idea is that when a pair of ring oscillators is selected, the voltage configuration of the ring oscillators is chosen so that the pair of ROs is guaranteed to work reliably across the whole temperature range.

The hardware architecture of our temperature-resistant PUF is shown in Figure 10. Just as for the original RO-PUF, our temperature-aware RO-PUF provides a maximal number of challenge/response pairs equal to $\frac{R(R-1)}{2}$, i.e. the challenge consists in selecting a pair of ring oscillators and the response is determined by comparing their oscillating frequencies. In our design, when a pair is selected, a voltage configuration is read from a memory and used to supply the inverters composing the selected ring oscillators. each pair is associated to a voltage configuration that guarantees reliable operation of the pair across the whole temperature range. Reliable configurations for each pair are pre-computed during the post-manufacturing testing of the PUF and stored in the memory.

As shown in Table II, for a UMC 90 nm PUF with $L = 2$, $C = 3$, each pair of ring oscillators can operate with $L^C = 8$ different voltage configurations. While some them are unreliable, some of these configurations will guarantee reliable operation across the whole temperature range from $-25°$ to $125°$. It is most probable to find for each pair at least one voltage configuration out of $L^C$ cases that is resistant to temperature variations across the whole temperature range. For example, as shown in Table II, for a given pair ... configurations among 8 are resistant to temperature variations. We tested this assumption for 100 pairs of ring oscillators; for each pair, we were able to find at least one case showing resistance to temperature variations.

A memory should put in correspondence each ring oscillator pair with a voltage configuration (see Table III). With a basic implementation, the size of the memory in terms of bits can be estimated by:

$$MEM_{bits} = \left\lceil \log_2^L \right\rceil \times C \times \frac{R \times (R-1)}{2}$$

where $L$ is the number of supply voltages, $C$ is the number of columns and $R$ is number of ring oscillators. As an example, for a PUF with $C = 5$, $L = 2$ and $R = 4$, the size of this memory will be 30 bits. It would be possible to introduce a controller and obtain memory savings exploiting the fact that the majority of the ring oscillator pairs is normally not temperature-sensitive and can operate reliably using the typical voltage configuration, but this requires further investigation and is outside the scope of this paper.

Since the entries are saved in a memory, they are potentially vulnerable to invasive and semi-invasive attacks. However,

| $v_1v_2v_3$ | 000 | | | 001 | | | 010 | | | 011 | | | 100 | | | 101 | | | 110 | | | 111 | | |
|---|---|---|---|---|---|---|---|---|---|---|---|---|---|---|---|---|---|---|---|---|---|---|---|---|
| temp. | t1 | t2 | t3 | t1 | t2 | t3 | t1 | t2 | t3 | t1 | t2 | t3 | t1 | t2 | t3 | t1 | t2 | t3 | t1 | t2 | t3 | t1 | t2 | t3 |
| PUF | 0 | 0 | 0 | 0 | 0 | 0 | 0 | 0 | 1 | 0 | 1 | 1 | 1 | 1 | 1 | 1 | 0 | 0 | 1 | 1 | 0 | 1 | 1 | 1 |

TABLE II
ONE PAIR OF RING OSCILLATORS WITH $L=2$ AND $C=3$ CAN OPERATE UNDER $L^C = 8$ VOLTAGE CONFIGURATIONS. THE EFFECT OF TEMPERATURE VARIATIONS FROM $-25°$ TO $125°$ FOR ALL THESE 8 CONFIGURATIONS ARE SHOWN.

| pairs | 1,2 | 1,3 | 1,4 | 2,3 | 2,4 | 3,4 |
|---|---|---|---|---|---|---|
| conf. | 01001 | 01100 | 10010 | 01110 | 11000 | 00110 |

TABLE III
SAMPLE MEMORY USED IN A TEMPERATURE-AWARE IMPLEMENTATION OF A RO-PUF WITH $L=2$, $C=5$ AND $R=4$. 1,2 MEANS THE PAIR BETWEEN RO1 AND RO2.

revealing the information in the memory does not give any extra information regarding the frequency of the two ROs in a pair and the RO-PUF has the same security as the solutions presented in [9] and [10].

Compared to [9], the memory is smaller in our design and a $100\%$ hardware utilization is obtained (each RO pair generates one output bit). Moreover, compared to both [9] and [10], our solution does not require the presence of any temperature sensor.

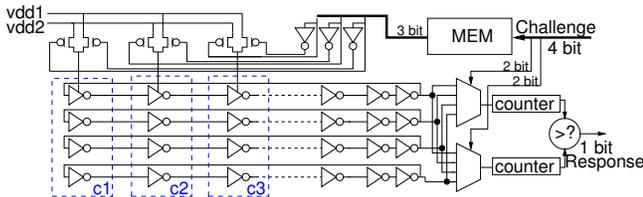

Fig. 10. Temperature-aware RO-PUF.

## IX. CONCLUSION AND FUTURE WORKS

In conclusion, our RO-PUF can support a high number of challenge/response pairs without impacting excessively the area of the PUF. It can also be used to implement a temperature-aware RO-PUF with a $100\%$ hardware utilization.

Issues that remain open and left to future work are a calculation of the number of valid challenge/response pairs and an efficient implementation of the temperature-aware RO-PUF. Also, the impact of supply voltage variations should investigated further.